\documentclass[prb,notitlepage,twocolumn]{revtex4}
\usepackage{color,xspace,graphicx}
\usepackage{amsmath}

\usepackage{lmodern}
\usepackage[caption=false]{subfig}

\renewcommand{\vec}[1]{\mathbf{#1}}

\newcommand{\sech}{\operatorname{\mathrm{sech}}}
\newcommand{\csch}{\operatorname{\mathrm{csch}}}

\newcommand{\realint}{\int d\vec r\,d t\,}
\newcommand{\fourint}{\int d\vec k\,d\omega\,}
\newcommand{\partint}{\int d\kpr\,d\omega}

\newcommand{\rll}{{r_{\parallel}}}
\newcommand{\rpr}{{r_{\perp}}}
\newcommand{\kll}{k_{\parallel}}
\newcommand{\kpr}{k_{\perp}}

\renewcommand{\deg}{\ensuremath{^{\circ}}\xspace}
\begin{document}
\title{Domain wall fluctuations in ferroelectrics coupled to strain}
\author{R.~T.~Brierley}\affiliation{TCM, Cavendish Laboratory, JJ Thomson Avenue, Cambridge, CB3 0HE, UK} \author{P.~B.~Littlewood} \affiliation{Physical Science and Engineering Division, Argonne National Laboratory, Argonne, IL 60439, USA}
\affiliation{James Franck Institute and Department of Physics, University of Chicago, IL 60637, USA}
\begin{abstract}
Using a Ginzburg--Landau--Devonshire model that includes  the coupling of polarization to strain, we calculate the fluctuation spectra of ferroelectric domain walls.
The influence of the strain coupling differs between 180\deg and 90\deg walls due to the different strain profiles of the two configurations.
The finite speed of acoustic phonons, $v_s$, retards the response of the strain to polarization fluctuations, and the results depend on $v_s$.
For $v_s\to\infty$, the strain mediates an instantaneous electrostrictive interaction, which is long-range in the 90\deg wall case.
For finite $v_s$, acoustic phonons damp the wall excitations, producing a continuum in the spectral function.
As $v_s\to0$, a gapped mode emerges, which corresponds to the polarization oscillating in a fixed strain potential.
\end{abstract}
\maketitle
\section{Introduction}
\label{sec:introduction}
Domain walls in ferroelectrics separate symmetry-equivalent directions of the order parameter, and are commonly present in bulk and thin films samples in order to minimize electrostatic and elastic energies.\cite{lines_principles_1977,catalan_domain_2012}
Because of the strong anisotropy of most ferroelectrics, the width of a domain wall is on the nm scale.\cite{catalan_domain_2012}
Since switching and domain dynamics involves wall motion, their dynamics plays an important role both in thermodynamic properties\cite{scott_electrocaloric_2011}, and in memory devices.\cite{scott_ferroelectric_1989}
As a result, understanding the behavior of domain walls is important for optimizing ferroelectrics for use in these contexts.

Studies of dynamics were initiated in $\text{PbTiO}_3$ by Merz \cite{merz_domain_1954}, who noted the unobservable width of domain walls in optical imaging, and the apparently activated kinetics.
The details of microscopic motion have only been resolved recently, with the guidance of {\it ab initio} theory and the connection to multi-scale dynamical models \cite{meyer_ab_2002,shin_nucleation_2007,zhang_nanodynamics_2011}.
However it was already noted by Merz that the velocity of domain walls did not appear to exceed the speed of sound; Dawber {\it et al.} \cite{dawber_perimeter_2003} noted that at small electric fields motion was limited by the viscous drag from acoustic emission; and direct imaging by piezoresponse microscopy \cite{gruverman_piezoresponse_2008} showed velocities not exceeding 100 m/s --- albeit at fields where the motion was still apparently activated.
Of course acoustic emission --- ferroelectric Barkhausen noise --- is long-known to be generated by domain wall motion.\cite{newton_observation_1949}

In a ferroelectric there is typically strong coupling of polarization to strain,\cite{lines_principles_1977} so the formation of a domain wall involves both electric and elastic degrees of freedom.
Both lead to long range polarization interactions.\cite{lines_principles_1977,catalan_domain_2012}
The elastic coupling is typically anisotropic, and means that ferroelectric domain walls form textures\cite{catalan_domain_2012} similar to those in ferroelastic materials such as martensites.\cite{bratkovsky_overview_1994,shenoy_martensitic_1999}
This restricts the domain wall orientations so that, for example, in two dimensions only 180\deg and 90\deg walls are allowed.\cite{rabe_physics_2007,hlinka_phenomenological_2006}
This introduces ferroelastic tweed structures, deviations from which are penalized by the long-range elastic forces.\cite{shenoy_martensitic_1999}
The finite speed of sound retards the strain-mediated interaction, and limits the speed of polarization switching, as already mentioned above.\cite{dawber_physics_2005,faran_twin_2010,catalan_domain_2012}

The dynamical behavior of domain walls is also significant in quantum phase transitions.\cite{sachdev_quantum_2011}
It has recently been suggested that unexpected behavior in the quantum phase transition of $\text{SrTiO}_3$ involves strain effects.\cite{rowley_quantum_2009,palova_quantum_2009}
Understanding the role of the strain dynamics in ferroelectric domain walls may help explain some of these results.

In this paper, we calculate the fluctuation absorption spectrum of a ferroelectric domain wall, including the effects of the anisotropic electrostrictive interactions.
Because our interest is in the clamping effects of strain and the consequent dynamics on long length scales, we use a coarse-grained model, the standard Ginzburg--Landau--Devonshire theory.\cite{rabe_physics_2007}
We will comment on the changes expected with the inclusion of atomicity.
Previous studies of ferroelectric domain walls using this approach have focused on quasi-static behavior.\cite{cao_theory_1991,marton_domain_2010,hlinka_phenomenological_2006}
The effects of domain wall fluctuations should be visible to experimental probes such as Raman spectroscopy\cite{stone_influence_2012,stone_local_2013} and inelastic neutron scattering.

We find that the long-wavelength spectrum can show two distinct modes: a low-frequency mode corresponding to motion of the domain wall; and a gapped mode due to fast oscillations of the polarization profile in the potential well of the slowly-responding strain.
The existence and approximate frequencies of these modes are in agreement with the {\it ab initio} calculations of Ref.~\onlinecite{zhang_nanodynamics_2011} for thin films, but our model makes the physical origin of the modes clear and describes the changes in behavior for different parameters.

We begin in Section \ref{sec:model} by summarizing our approximate model for the domain walls, which is based on a mean field theory for the polarization and elastic degrees of freedom.
Section \ref{sec:fluctuation-spectrum} describes how we calculate the domain wall fluctuation spectrum.
In Section \ref{sec:results}, we present our results for the spectral function of wall fluctuations, in the limits of infinite (Sec.~\ref{sec:inst-resp}) and finite sound speeds (Sec.~\ref{sec:retarded-response}).
In Section \ref{sec:conclusions} we present our conclusions.

\section{Preliminaries}
\label{sec:model}
In this section we present the simplified model we use to describe the polarization and elastic degrees of freedom, specified to pseudo-cubic systems for simplicity.
The treatment is similar to other discussions of domain walls using Ginzburg--Landau--Devonshire models.\cite{cao_landau-ginzburg_1990,cao_theory_1991,hlinka_phenomenological_2006,marton_domain_2010,rabe_physics_2007}
Throughout this paper, we use a two-dimensional model for simplicity.
However, the results we present would be qualitatively the same in higher dimensions.
As we wish to focus on the role that strain plays in the domain wall fluctuations, we assume that the polarization-strain coupling is the only source of anisotropy.

We use an action for the combined system with terms describing the polarization, elastic degrees of freedom, and long-range dipolar interactions separately.
\begin{gather}
  \label{eq:bitty-action}
  S=T_{\text{pol}}-V_\text{pol}-V_\text{dip}+T_\text{el}-V_\text{el}-V_\text{pol-el}\text,
\end{gather}
where $T_\text{pol}-V_\text{pol}$ and $T_\text{el}-V_\text{el}$ are the kinetic and potential energy contributions from the polarization and elastic degrees of freedom, $V_\text{dip}$ describes the long range part of the dipolar interactions and $V_\text{pol-el}$ gives the interaction between elastic and polarization degrees of freedom.

For the polarization, we use an action of the form:
\begin{gather}
  \label{eq:pol-action}
 T_\text{pol}=  \realint |\partial_t\vec P|^2 \\
 V_\text{pol}= \realint\left(\gamma (\nabla \vec P)^2+\alpha|\vec P|^2+\beta |\vec P|^4\right)\text,
\end{gather}
where $\vec P$ is the two-dimensional polarization vector and $\gamma$ is the gradient contribution to the domain wall energy.
The units of time are chosen so that there is no constant in Eq.~\eqref{eq:pol-action}.
We assume that $\alpha<0$ and $\beta$ is large enough that the transition is second order.
This simplifies the description of the model and the form of the domain wall, discussed below.
Many ferroelectric transitions of interest are first order\cite{rabe_physics_2007} and so violate this assumption but describing the details of the ferroelectric transition are not important for this paper.
The form of domain walls is similar in the case of first and second order transitions\cite{cao_landau-ginzburg_1990,cao_theory_1991} and so we expect the model presented here to capture the relevant physical phenomena.

We describe the long-range part of the dipolar interaction using: \cite{ahluwalia_piezoelectric_2004}
\begin{gather}
  \label{eq:dip-action}
    V_\text{dip}=\fourint\sum_{\alpha,\beta}P_\alpha^*(k)g\frac{k_\alpha k_\beta}{k^2} P_\beta(k)\text.
\end{gather}
$g$ is the strength of the interaction, which we take to be an isotropic constant for simplicity.

The elastic degrees of freedom are given by the atomic lattice positions $\vec u$.
The kinetic energy term is:
\begin{equation}
  \label{eq:ke-elastic}
  T_{el}=\realint \frac{a_2}{4v_s^2}|\dot{\vec{u}}|^2
\end{equation}
$v_s$ characterizes the speed of acoustic phonons and $a_2$ is one of the elastic constants, which appears in this form for later convenience.

The elastic potential energy is mostly clearly written in terms of strain fields:\cite{ahluwalia_piezoelectric_2004}
\begin{gather}
  \label{eq:strain-rep}
  V_\text{el}=\realint\left(\frac {a_1}2\phi_1^2+\frac {a_2}2\phi_2^2+\frac {a_3}2\phi_3^2\right).
\end{gather}
The strains $\phi_1=(\partial_xu_x+\partial_yu_y)/\sqrt2$, $\phi_2=(\partial_xu_x-\partial_yu_y)/\sqrt2$ and $\phi_3=(\partial_xu_y+\partial_yu_x)/2$ are subject to the compatibility condition:\cite{shenoy_martensitic_1999}
\begin{equation}
  \label{eq:compatibility}
  \nabla^2 \phi_1-(\partial_x^2-\partial_y^2)\phi_2-2\sqrt2\partial_x\partial_y\phi_3=0\text.
\end{equation}

We assume that the polarization couples only to one strain field, $\phi_2$.
This corresponds to the formation of the polarization being accompanied by a distortion of the local lattice from square to rectangular.
The coupling takes the form:
\begin{equation}
  \label{eq:strain-pol-coup}
    V_\text{pol-el}=\realint q \phi_2(P_x^2-P_y^2)
\end{equation}
where $q$ is the strength of the strain interaction.

When $a_2=a_3/2$, the system is elastically isotropic and the shear strains $\phi_2$ and $\phi_3$ become equivalent.
However, the anisotropic form of the coupling, Eq.~\eqref{eq:strain-pol-coup}, means the model discussed here never has complete rotational symmetry.
The behavior for different values of $a_2$ is not qualitatively different.

As the strain terms are quadratic, we can eliminate the strain degrees of freedom, yielding a description of the interactions in terms of a non-local kernel for electrostriction:\cite{ahluwalia_influence_2000,lookman_ferroelastic_2003}
\begin{widetext}
\begin{align}
  \label{eq:p-only-action}
V_\text{pol-el}+V_\text{el}&\to V_\text{str} = -\eta\fourint|\Gamma(\vec k,\omega)|^2 H(\vec k)\text,\\
H(\vec k,\omega)&=
\begin{cases}
  1&\vec k=\vec 0\text,\\
\frac{(A+B-1)(k_x^2+k_y^2)^2-4(B-1)k_x^2k_y^2-C\Omega^2(k_x^2+k_y^2)}{A(k_x^2+k_y^2)^2+B(k_x^2-k_y^2)^2+(C+A)\Omega^2(k_x^2+k_y^2)-C\Omega^4}&\text{otherwise,}
\end{cases}
\\
\Gamma(\vec k,\omega)&=\int d\vec r e^{-i\vec k\cdot \vec r}(P_x(\vec r,t)^2-P_y(\vec r,t)^2)\text,\label{eq:gamma-def}
\end{align}
\end{widetext}
where $\Omega=\omega/v_s$, the elastic constants are written in terms of $A=\frac{a_2}{a_1}+2\frac{a_2}{a_3}$, $B=1-2\frac{a_2}{a_3}$ and $C=\frac{2a_2^2}{a_1a_3}=(A+B-1)(1-B)$; and $\eta=q^2/2a_2$ is the electrostrictive constant.

The conditional value of $H(\vec k)$ arises from the separate treatment of homogeneous strains $\phi_i(\vec k=0)$ as they do not need to satisfy the compatibility condition.\cite{larkin_phase_1969-1}
$H(\vec k)$ is non-analytic and direction-dependent at at the origin, due to the long range and anisotropic nature of the elastic interactions.
Static textures for which $k_x=\pm k_y$ also have $H(\vec k)=1$, as they satisfy the compatibility condition while keeping $\phi_1=\phi_3=0$.
This produces the tweed structure of ferroelastic and 90\deg ferroelectric domains.\cite{bratkovsky_overview_1994,shenoy_martensitic_1999}

The poles of $H(\vec k,\omega)$ are the acoustic phonon frequencies.
When $B=0$, they are given by $\omega=v_s|\vec k|$, corresponding to the transverse phonons and $\omega=\sqrt{(a_1+a_2)/a_2}v_s|\vec k|$ for the longitudinal phonons.

\subsection{Mean field solution and static domain wall}
\label{sec:appr-wall-form}
In this subsection, we summarize the static mean-field solution to our model, Eq.~\eqref{eq:bitty-action} and describe the approximate form of the domain wall we use in the rest of the paper.
Due to the anisotropic strain coupling, the system has only discrete symmetry.
In the mean-field approximation, the polarization, $\vec P$, in equilibrium is:
\begin{gather}
  \label{eq:mf-polarization}
  \vec P=\begin{cases}
    \vec 0&\text{if $\alpha>0$,}\\
    \begin{pmatrix}
      \pm\rho\\0
    \end{pmatrix}\text{ or }
    \begin{pmatrix}
      0\\\pm\rho
    \end{pmatrix}&\text{if $\alpha<0$},
  \end{cases}\\
\text{where}\quad \rho^2=\frac{-r}{2(\beta-\eta)}\text.
\end{gather}
We have assumed that ${\beta-\eta}>0$, giving a second order transition.

We consider a single domain wall interpolating between the different polarization orientations of Eq.~\eqref{eq:mf-polarization}.
If a wall is charged, then it incurs an extensive energy cost due to the long-range dipolar forces, Eq.~\eqref{eq:dip-action}.
This is avoided if $P_\perp$ is the same on each side of the domain wall, so that $\nabla\cdot\vec P=0$ and the wall is neutral.
We use an approximate description for the form of a domain wall:
\begin{equation}
  \label{eq:dw-prototype}
  \vec P=P_\perp+P_\parallel \tanh\left(\frac{\rpr-f(\rll,t)}L\right)\text.
\end{equation}
The directions $\rpr$ and $\rll$ are perpendicular and parallel to the domain wall, respectively.
$f(\rll,t)$ is a field describing the fluctuations of the wall, so $f(\rll,t)=0$ is the minimum energy configuration.
This form is similar to more accurately calculated numerical results. \cite{marton_domain_2010,hlinka_phenomenological_2006}

\begin{figure}
  \centering
  \subfloat[180\deg wall]{\includegraphics[width=0.2\textwidth]{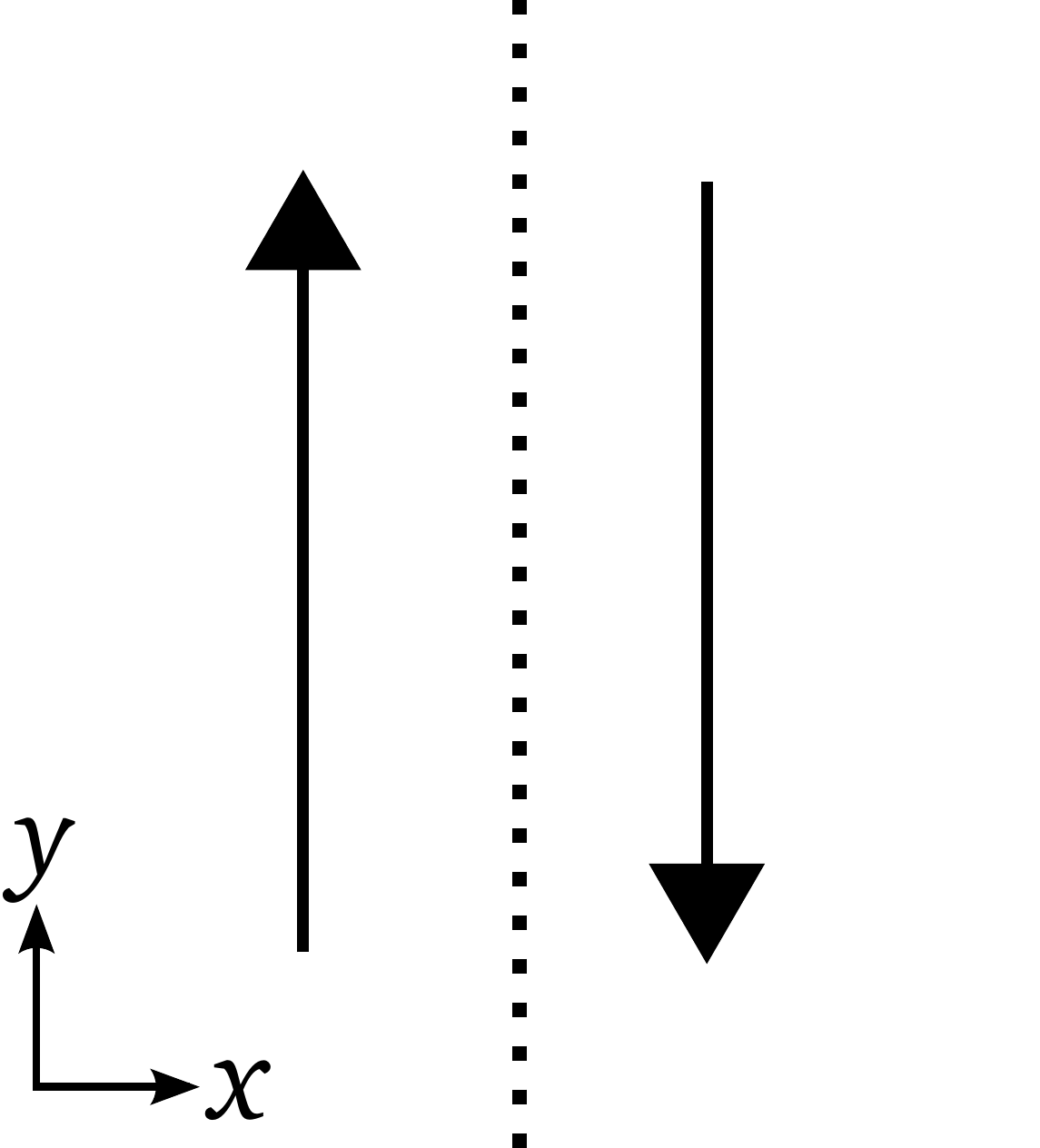}}
  \subfloat[90\deg wall]{\includegraphics[width=0.2\textwidth]{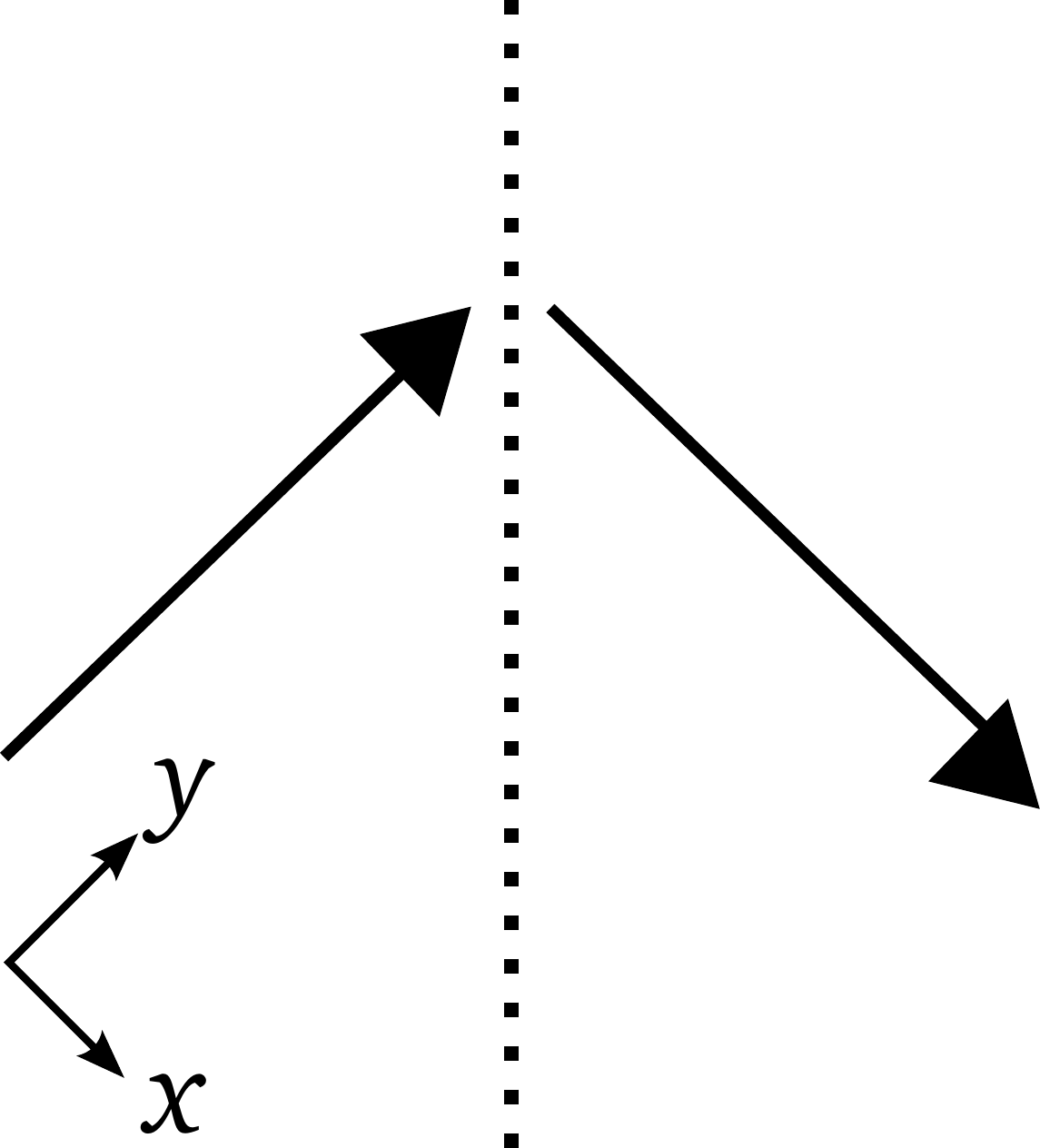}}
  \caption{Examples of the two configurations allowed in the two-dimensional model ferroelectric due to the combination of dipolar and strain interactions.}
  \label{fig:doma-wall-conf}
\end{figure}

The combination of the mean field orientations, Eq.~\eqref{eq:mf-polarization}, with the dipole interactions, Eq.~\eqref{eq:dip-action}, mean there are only two domain wall configurations, Fig.~\ref{fig:doma-wall-conf}.

A 180\deg wall runs parallel to the polarization so that $P_\perp=0$.
We take $\rpr=x$ and $\rll=y$ for the 180\deg wall.
Then:
\begin{gather}
  \label{eq:180-wall}
  \vec P_\text{180}
=\rho
  \begin{pmatrix}
    0\\\tanh\left(\frac{\rpr-f(\rll,t)}{L_\text{180}}\right)
  \end{pmatrix}\text,\\
  \phi_{\text{2,180}}=-q\rho^2\left[\frac{1}{a_2}-\frac{\sech^2(\rpr/L_\text{180})}{a_1+a_2}\right]\text,\\
  \phi_{\text{1,180}}=q\rho^2\frac{\sech^2(\rpr/L_{\text{180}})}{a_1+a_2}\label{eq:180-wall-phi1}\text,\\
  \phi_{\text{3,180}}=0\text.\label{eq:180-wall-phi2}
\end{gather}

A 90\deg domain wall is oriented at 45\deg to the polarization, with $P_\perp=\rho/\sqrt2$ so that:
\begin{gather}
  \label{eq:90-wall}
  \vec P_{\text{90}} =\frac1{2}\left[
  \begin{pmatrix}
    \rho\\-\rho
  \end{pmatrix}
  +
  \begin{pmatrix}
    \rho\\
\rho
  \end{pmatrix}\tanh\left(\frac{ \rpr-f(\rll,t)}{L_\text{90}}\right)
\right]\\
\phi_{2,\text{90}}=-q/a_2\rho^2\tanh(\rpr/L_\text{90})\label{eq:90-wall-phi2}
\end{gather}
where $\rll = (x\pm y) / \sqrt{2}$, $\rpr = (x\mp y) / \sqrt{2}$.
Here $\phi_1=\phi_3=0$ because the 90\deg wall satisfies the compatibility condition and forms a ferroelastic twin boundary.\cite{cao_landau-ginzburg_1990}

The wall widths can be estimated variationally, giving:
\begin{gather}
  \label{eq:wall-widths}
L^2_\text{180}=\frac{\gamma}{\rho^2\left(\beta-\eta\frac{a_2}{a_1+a_2}\right)}\text,\\
L^2_\text{90}=\frac{\gamma}{\rho^2\left(\beta-\eta\right)}\text.
\end{gather}

The atomic positions $\vec u \sim \int \phi_2 d\rpr$, so the influence of a domain wall on $\vec u$ depends on the inhomogeneous part of $\phi_2$, which has a different form for 180\deg and 90\deg walls.
In the 180\deg case, it goes as $\sech\rpr/L$, and so is localized in the region of the wall.
As a result, moving the wall does not have an effect on the atomic positions a long way from the wall center.
In contrast, a 90\deg wall has inhomogeneous strain that goes as $\tanh\rpr/L$.
This means that changing the wall position involves changing the atom displacement $\vec u$ throughout the crystal,\cite{barsch_dynamics_1987,horovitz_twin_1991} and so strain effects will be longer range in the 90\deg case.

\section{Fluctuation action}
\label{sec:fluctuation-spectrum}
To calculate the fluctuation spectrum of the wall, we assume that fluctuations away from the equilibrium configuration, $f(\rll,t)=0$ and expand to second order.
This produces a fluctuation action of the form:
\begin{gather}
  \label{eq:schematic-fluc-action}
  V_\text{fluc}=\int d\omega\,d\kll\, G^{-1}(\kll,\omega)|f(\kll,\omega)|^2\text.
\end{gather}
Under linear response, \cite{chaikin_principles_1995} $A(\omega,\kll)=\Im G(\omega+i0,\kll)$ gives the absorption spectrum of the system.
The poles of $G(\omega,\kll)$ give the normal mode frequencies, and these appear as peaks in $A(\omega,\kll)$, with broadening in the case of a decaying mode.

By the construction of Eq.~\eqref{eq:dw-prototype}, only  $P_\parallel$ changes, and only in the $\kll$ direction.
As $\rho$ and $L$ have been determined variationally, there is no contribution from the non-gradient terms of $V_\text{pol}$ or any parts of the action that depend only on $\kpr$.

The fluctuation of the polarization field is:
\begin{gather}
  \label{eq:pol-expansion-90}
  \begin{aligned}
  P_{\parallel}(\vec k,\omega)&=\bar\rho\int d\rpr\,\tanh\frac{\rpr}{L}e^{-i\kpr(\rpr-f(\rll,t))-i\kll \rll+i\omega t}\\
  &\approx\frac{\bar\rho L}2i\csch\frac{\pi \kpr L}2\left(\delta_{\kll}-i\kpr f(\kll,\omega)+\dots\right)\text,
\end{aligned}
\end{gather}
where $\bar\rho=\rho$ for a 180\deg wall and $\bar\rho=\rho/\sqrt{2}$ for a 90\deg wall, c.f.\ Eqs.~\eqref{eq:180-wall} and \eqref{eq:90-wall}.
\begin{widetext}
Substituting Eq.~\eqref{eq:pol-expansion-90} into the gradient term, Eq.~\eqref{eq:pol-action}, makes the contribution:
\begin{gather}
  \label{eq:fluc-grad}
  \begin{aligned}
  \delta (V_\text{pol}-T_\text{pol})&=\fourint(-\omega^2+\gamma\kll^2)|f(\kll,\omega)|^2\frac{\bar\rho^2 L^2}4\kpr^2\csch^2\frac{\pi\kpr L}2\\
  &=\partint\frac{2\bar\rho^2}{3\pi L}(-\omega^2+\gamma\kll^2)|f(\kll,\omega)|^2\text,
\end{aligned}
\end{gather}
while the long-range dipolar term, Eq.~\eqref{eq:dip-action}, gives:
\begin{gather}
  \label{eq:dip-expansion}
  \begin{aligned}
    \delta V_\text{dip}&=-g\fourint\frac{\kll\kll}{\kll^2+\kpr^2} P_\parallel P_\parallel\\
    &=-\frac{g\bar\rho^2 L^2}4\fourint\frac{\kll^2}{\kll^2+\kpr^2}\kpr^2\csch^2\frac{\pi\kpr L}2 |f(\kll,\omega)|^2\text.
  \end{aligned}
\end{gather}
Finally, the strain contribution involves the inhomogeneous part of $\Gamma(\vec k,\omega)$, Eq.~\eqref{eq:gamma-def}:
\begin{gather}
  \label{eq:gamma-90}
  \begin{aligned}
    \Gamma(\vec k,\omega)&=\int d\rll\,d\omega\,\Gamma_0(\kpr) e^{i(\omega t- \kpr
      f(\rll,t))}\\&\approx\Gamma_0(\kpr)\left(\delta_{\kll,\omega}-i\kpr
      f(\omega,\kll)-\frac{\kpr^2}2\sum_{\omega',q}f(\kll+q,\omega+\omega')f(-q,-\omega')\right)\text,
  \end{aligned}
\end{gather}
where the form in the absence of fluctuations, $\Gamma_0(\kpr)$, is:
\begin{gather}
  \label{eq:gamma-cases}
  \Gamma_0(\kpr)=
  \begin{cases}
    i\frac{\rho^2L}2\csch\frac{\pi\kpr L}2&\text{90\deg wall,}\\
    -\frac{\rho^2L}2\kpr L\csch\frac{\pi\kpr L}2&\text{180\deg wall.}
  \end{cases}
\end{gather}
The extra power of $\kpr$ for the 180\deg wall reflects the stronger localization of the wall's strain profile in this case, as discussed at the end of Sec.~\ref{sec:appr-wall-form}. 
Placing this expression in Eq.~\eqref{eq:p-only-action} produces the fluctuation term:
\begin{gather}
  \label{eq:strain-expansion}
  \begin{aligned}
    \delta V_{\text{str}}&=-\fourint\eta|\Gamma_0(\kpr)|^2H(\kpr,\kll,\omega)\left(\kpr^2|f(\kll,\omega)|^2-\delta_\omega\delta_{\kll}\kpr^2\int dq\,d\omega'\,f(\kll+q,\omega+\omega')f(-q,-\omega')\right)\\
    &=-\fourint\eta|\Gamma_0(\kpr)|^2\kpr^2\left(H(\kpr,\kll,\omega)-H(\kpr,0,0)\right)|f(\omega,\kll)|^2\text.
  \end{aligned}
\end{gather}

In the 180\deg case, the strain contribution to the fluctuation action is then:
  \begin{gather}
    \label{eq:fluc-strain-180}
    \delta V_\text{str,180}=-\frac{\eta}4\frac{\rho^4L^4}{A+B}
    \fourint\csch^2\frac{\pi\kpr
      L}2\kpr^4\frac{C(A+B-1)\Omega^4+4A\kll^2\kpr^2+(A+B-1)^2\Omega^2(\kpr^2+\kll^2)}{A(\kpr^2+\kll^2)^2+B(\kpr^2-\kll^2)^2+(C+A)\Omega^2(\kpr^2+\kll^2)-C\Omega^4}|f(\kll,\omega)|^2\text.
  \end{gather}
For a 90\deg wall $H(\kpr,\kll,\omega)$ is Eq.~\eqref{eq:strain-rep} with the coordinates rotated by 45\deg to match the wall axes $\rpr$ and $\rll$  (see Fig.~\ref{fig:doma-wall-conf}).
Substituting into Eq.~\eqref{eq:strain-expansion} gives:
\begin{gather}
  \label{eq:fluc-strain-90}
  \delta V_\text{str,90}=-\frac{\eta}{4}\rho^4L^2\fourint \csch^2\frac{\pi\kpr L}2\kpr^2\frac{(C\Omega^4+A\Omega^2(\kpr^2+\kll^2)-4\kpr^2\kll^2)}{A(\kpr^2+\kll^2)^2+B\kpr^2\kll^2-C\Omega^4+\Omega^2
    (\kpr^2+\kll^2)(A+C)}|f(\kll,\omega)|^2\text.
\end{gather}
\end{widetext}

\section{Results}
\label{sec:results}

\subsection{Instantaneous response}
\label{sec:inst-resp}
We begin by considering the $v_s\to\infty$ limit, so that $\Omega\to0$.
In this limit, the phonons mediate an instantaneous electrostrictive interaction.
The strain part of the fluctuation action simplifies to
\begin{widetext}
  \begin{gather}
    \label{eq:fluc-strain-stat-180}
    \delta V_{\text{str,180}}=-\fourint \frac \eta4\frac{\rho^4L^4}{A+B}\frac{4A\kll^2\kpr^2}{B(\kpr^2-\kll^2)^2+A(\kpr^2+\kll^2)^2}\kpr^4\csch^2\frac{\pi\kpr L}2\\
    \delta V_{\text{str,90}}=\fourint
    \frac \eta{4}\rho^4L^2\frac{4\kll^2\kpr^2}{4B\kpr^2\kll^2+A(\kpr^2+\kll^2)^2}\kpr^2\csch^2\frac{\pi\kpr
      L}2\label{eq:fluc-strain-stat-90}
  \end{gather}
\end{widetext}
Note that $\delta V_{\text{str,180}}$ gives a negative contribution to the fluctuation energy.
The strain interactions favor a domain wall that is oriented at 45\deg to the polarization, as in the 90\deg wall.
This forms a ferroelastic twin\cite{cao_landau-ginzburg_1990} and would automatically satisfy the compatibility relation Eq.~\eqref{eq:compatibility} without requiring non-zero $\phi_1$ and $\phi_3$ [Eqs.~\eqref{eq:180-wall-phi1} and \eqref{eq:180-wall-phi2}].
The orientation of a 180\deg wall is instead stabilized by the dipolar forces and wall fluctuations are penalized by the other contributions Eqs.~\eqref{eq:fluc-grad} and \eqref{eq:dip-expansion}.
If $g$ and $\gamma$ were too small compared to $\eta$ there would be an instability at finite wavevector.
We assume here that they are large enough that this is not an issue.

As $\Omega\to0$, the frequency only appears in the contribution of
Eq.~\eqref{eq:pol-expansion-90}, so the poles of $G$ are given by a
dispersion relation of the form:
\begin{gather}
  \label{eq:dispersion-expression}
  \omega^2=\gamma \kll^2+\frac{3\pi L}{2\bar\rho^2}\left[g_\text{dip}(k)+g_\text{str}(k)\right]
\end{gather}
where the $g(\kll)$ are the integrands of Eqs.~\eqref{eq:fluc-strain-stat-180} and \eqref{eq:fluc-strain-stat-90} integrated over $\kpr$.
In the limit of small wavevectors, $\kll L\ll 1$, these integrals can be approximated, giving, in the 180\deg case:

\begin{gather}
  \label{eq:dispersion-180}
  \omega^2=\kll L\left(\frac32 g\right)+\kll^2L^2\left(-\frac{4A\eta\rho^2}{(A+B)^2}-\frac32 g+\frac{\gamma}{L^2}\right)+\dots
\end{gather}
and in the 90\deg case:
\begin{gather}
  \label{eq:dispersion-90}
  \omega^2=\kll L\left(\frac32 g+\frac{12\eta\rho^2}\pi c\right)+\kll^2L^2\left(\frac{\gamma}{L^2}-\frac32 g-12\eta\rho^2\right)+\dots
\end{gather}
where $c$ is a constant:
\begin{gather}
  \label{eq:const-c-integral}
  c=\int^\infty_{-\infty}du\,\frac{u^2}{A(u^2+1)^2+4Bu^2}\text.
\end{gather}
In the isotropic case, $B=0$ and $c=\pi/2A$. The different ranges of the electrostrictive effects, discussed at the end of Sec.~\ref{sec:appr-wall-form} are reflected in the appearance of strain terms at order $\kll^2$ in Eq.~\eqref{eq:dispersion-180} but $\kll$ in Eq.~\eqref{eq:dispersion-90}.

At large wavevectors, $\kll L \gg 1$, the finite size of the wall leads to an exponential cutoff from the $\csch\pi k_x L/2$ part of $\Gamma_0(\kpr)$ and the dispersion tends to $\omega\sim\sqrt\gamma k$.

\subsection{Retarded response}
\label{sec:retarded-response}
\begin{figure*}
  \centering
\subfloat[$v_s/L=\frac13\Delta_{90}$]{\includegraphics[width=0.3\textwidth]{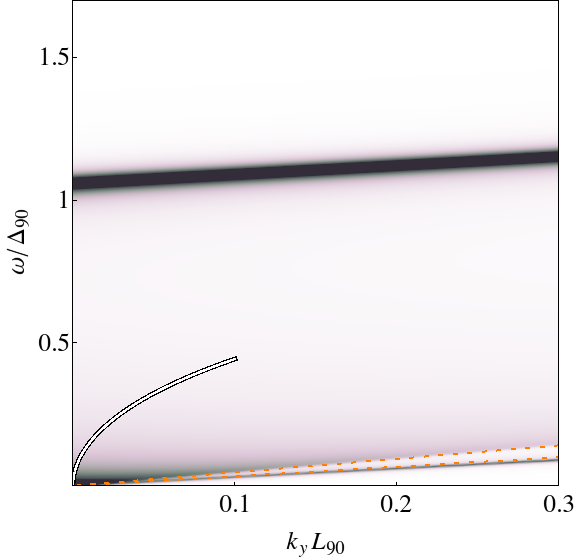}\label{fig:90-0.3}\label{fig:90-gapped}}
\subfloat[$v_s/L=\Delta_{90}$]{\includegraphics[width=0.3\textwidth]{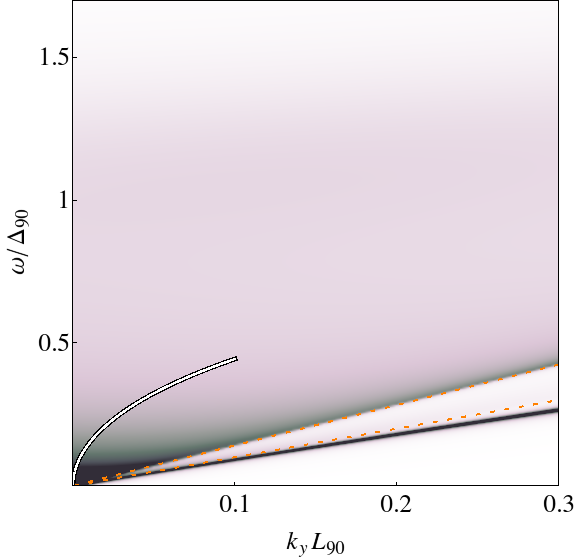}}
\subfloat[$v_s/L=3\Delta_{90}$]{\includegraphics[width=0.3\textwidth]{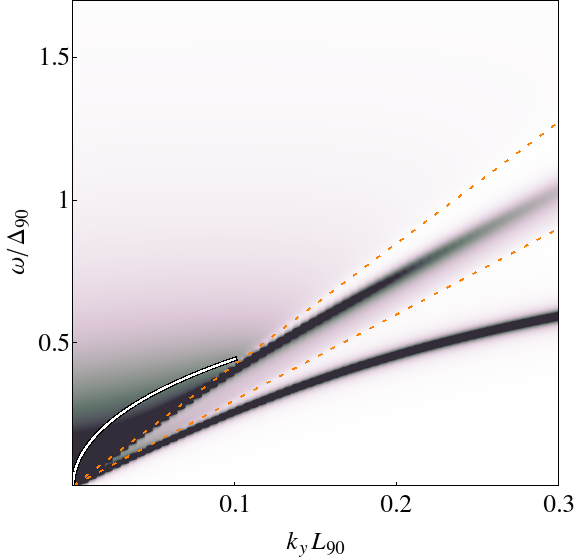}}
  \caption{Spectral functions $A(\kll,\omega)=\Im G(\kll,\omega+i0)$ in arbitrary units for 90\deg domain walls with (a) $v_s/L=\frac{1}{3}\Delta_{90}$, (b) $\Delta_{90}$ and (c) $3\Delta_{90}$. $\Delta_{90}$ is the $\kll$ frequency of the gapped mode when $v_s\to0$. The color scale is saturated and the same scaling is used for all figures.  $g=1$, $\gamma=1$, $\eta=1$ for simplicity.  Solid line (white and black): For comparison, the dispersion curve for small $\kll$ when  $v_s\to \infty$ and the strain interaction becomes instantaneous, Eq.~\eqref{eq:dispersion-90}. Dashed lines (orange):  $\omega=v_{l,t}\kll$, the threshold frequencies for decay into longitudinal and transverse acoustic phonons.}
  \label{fig:spectr-funct-90deg}
\end{figure*}

\begin{figure*}
  \centering
\subfloat[$v_s/L=\frac13\Delta_{180}$]{\includegraphics[width=0.3\textwidth]{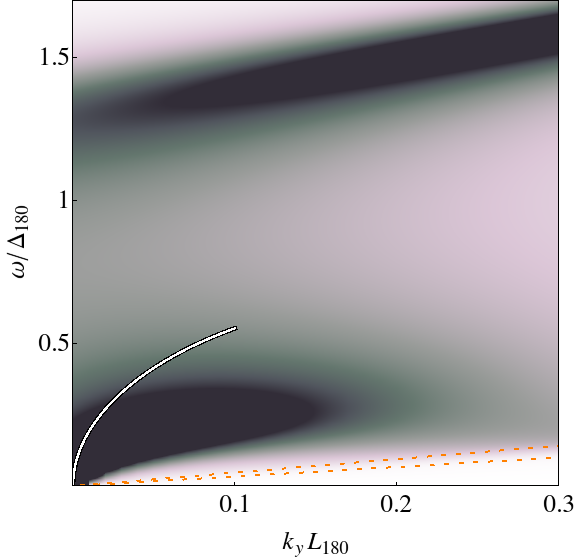}\label{fig:180-gapped}}
\subfloat[$v_s/L=\Delta_{180}$]{\includegraphics[width=0.3\textwidth]{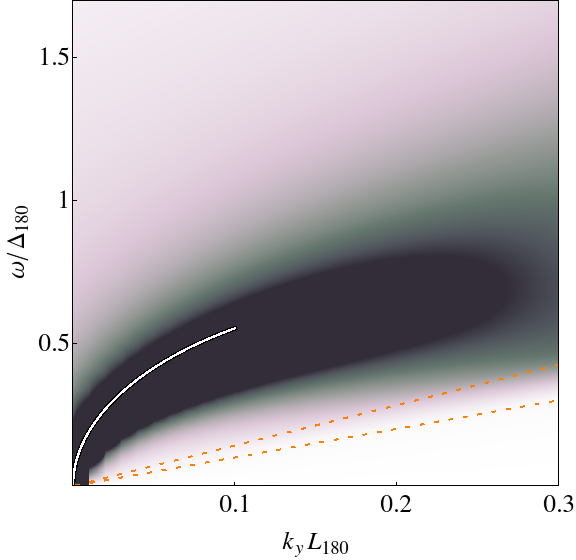}}
\subfloat[$v_s/L=3\Delta_{180}$]{\includegraphics[width=0.3\textwidth]{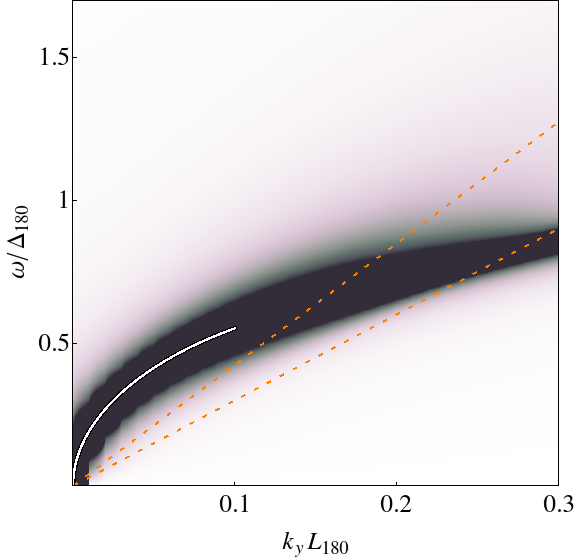}\label{fig:180-fast}}
  \caption{As in Fig.~\ref{fig:spectr-funct-90deg} for 180\deg domain walls. The $v_s\to\infty$ dispersion, white and black line, is given by Eq.~\eqref{eq:dispersion-180}.}
  \label{fig:spectr-funct-180deg}
\end{figure*}

Away from the limit $v_s\to\infty$, the response of the strain to changes in the polarization is no longer instantaneous and so the strain mediated interaction is retarded.
Examples of the spectral functions $A(\omega,\kll)$ are shown in Figs.~\ref{fig:spectr-funct-90deg} and \ref{fig:spectr-funct-180deg}.
We have used $g=1$, $\gamma/L^2=1$, $A=2$, $B=0$, $\eta=1$ for illustrative purposes, as this shows the qualitative behavior, and these quantities are often a similar order of magnitude.\cite{hlinka_phenomenological_2006,rabe_physics_2007}
We now discuss the origin of the features shown in Figs.~\ref{fig:spectr-funct-90deg} and \ref{fig:spectr-funct-180deg}.

In the limit  $v_s\to0$, the strain does not change in response to fluctuations of the polarization.
This means that the center of the polarization domain wall moves in a static potential caused by the strain.
These polarization oscillations therefore have a finite frequency, determined by the mean field strain, which depends on the mean field polarization. 
The frequency gap is $\Delta^2_\text{90}=2\eta \rho^2$ for the 90\deg wall and $\Delta^2_\text{180}=\eta\rho^2\frac45\frac{A+B-1}{A+B}$ in the 180\deg case.
This gapped mode is visible in Figs.~\ref{fig:90-gapped} and \ref{fig:180-gapped}.


The strain interaction arises from the coupling of wall oscillations, which only have momentum in the $\kll$ direction, to acoustic phonons that may have a $\kpr$ component.
Phonons can therefore scatter from the domain wall by creating wall excitations and change their momentum in the $\kpr$ direction, as observed in recent Raman scattering experiments.\cite{stone_influence_2012}
Equally, a wall excitation may decay into a continuum of phonons with different values of $\kpr$ as long as $\kll$ and the excitation energy $\omega$ are conserved.
This decay process damps the wall oscillations and produces broadened regions in the absorption spectrum.
For a given $\kll$ there is a minimum $\omega$ required to create longitudinal or transverse acoustic phonons.
This leads to thresholds, plotted in Figs.~\ref{fig:spectr-funct-90deg} and \ref{fig:spectr-funct-180deg}, above which decay into the phonon branches becomes possible and the response is broadened.

The fluctuation damping depends on the strength of coupling to phonons with the same $\kll$ and the appropriate $\kpr$ so that they have the same frequency, $\omega \sim v_s|\vec k|$.
The coupling is given by the behavior of the integrands in Eqs.~\eqref{eq:fluc-strain-180} and \eqref{eq:fluc-strain-90}.
The finite size of the wall, $L$, leads to an exponential cutoff at large wavevectors $\kpr\gtrsim 1/L$.
The damping therefore cuts off for frequencies $\omega > \omega_c\sim v_s/L$.

The coupling to phonons vanishes as $\kll\to0$ for a 180\deg wall, but tends to a constant in the 90\deg case.
This is a result of the coupling of a  90\deg wall to long-range atomic motion, while the strain effects in a 180\deg wall are more localized, as discussed in Sec.~\ref{sec:appr-wall-form}.
This means that, for sufficiently low $\omega$, the 180\deg wall is only coupled to relatively fast phonon modes with finite $\kpr$, and the spectrum shows the $\sqrt\kll$ behavior of the instantaneous response, see Fig.~\ref{fig:spectr-funct-180deg}.
In contrast, motion of the 90\deg wall couples to phonons with arbitrarily long wavelength, and so it is limited by the phonon velocity at small $\omega$.
This leads to a mode that follows $v_s\kll$, as is visible in Fig.~\ref{fig:spectr-funct-90deg}.

There is always a peak in the spectral density as $\omega\to0$, $\kll\to0$. This is required as our model is translationally invariant so that a uniform $\kll = 0$ shift of the domain wall should not cost any energy.

The gapped mode may lie within the region where the wall excitations are damped by the acoustic phonons.
The gapped mode is then damped and broadened so that it is  eventually no longer a distinct feature.
As shown in Figs.~\ref{fig:spectr-funct-90deg} and \ref{fig:spectr-funct-180deg}, the gapped mode is no longer visible if $\Delta\sim\sqrt{\eta\rho^2}\lesssim v_s/L$.
These quantities are both around $10$ {meV} in several ferroelectric materials\cite{hlinka_phenomenological_2006,rabe_physics_2007,bell_elastic_1963,faran_twin_2010,zhang_nanodynamics_2011}
 and so an analysis of the peak visibility in a given material will depend on details beyond those considered here.

\section{Conclusions}
\label{sec:conclusions}

We have calculated the spectral functions for fluctuations of both 180\deg and 90\deg ferroelectric domain walls, including the effects of strain coupling.
The speed of the acoustic phonons, $v_s$, determines how quickly the strain can respond to changes in the polarization and so retards the strain-mediated electrostrictive interaction.

If $v_s\to\infty$, the electrostrictive interaction is instantaneous and the wall fluctuations satisfy a dispersion relation, Eq.~\eqref{eq:dispersion-expression}.
In both the 180\deg and 90\deg wall cases, dipolar forces introduce a long range interaction, which gives a $\sqrt \kll$ dispersion at small wavevectors.
90\deg walls couple to atomic displacements a long distance from the wall,\cite{barsch_dynamics_1987} and so the strain acts as a long range interaction, contributing an additional $\sqrt\kll$ term, Eq.~\eqref{eq:dispersion-90}.
For the 180\deg wall, the strain interaction is shorter range as the strain variation is more localized.
The first strain contribution to the dispersion is then of order $\kll$.

When $v_s$ is finite, the electrostrictive interaction is retarded and domain wall fluctuations are damped by emission of acoustic phonons.
This produces a continuum response in the spectral function.
The finite size of the domain wall cuts off the coupling to acoustic phonons at frequencies $\omega_c\sim v_s/L$.
It also introduces an additional mode of oscillation, which in the limit $v_s\to 0$ corresponds to the wall moving in a static strain potential.
This mode is gapped, with $\Delta\sim \sqrt{\eta\rho^2}$.
However, if $\Delta<\omega_c$, it will be damped by the acoustic phonon continuum and so may be broadened and indistinct to experimental probes such as Raman spectroscopy\cite{stone_influence_2012,stone_local_2013} or neutron scattering.
The {\it ab initio} results of Ref.~\onlinecite{zhang_nanodynamics_2011} also show a gapped mode for domain wall fluctuations, with the appropriate frequency, $\sim 10$ meV.
Our Ginzburg--Landau--Devonshire analysis makes the physics of this new mode transparent, and demonstrates its behavior for different material parameters. 

Finally we remark on the influence of physics on the unit cell scale, neglected in our analysis.
This breaks translational invariance, hence introducing a gap determined by the ratio of the domain wall width to the unit cell.
A domain wall is then described more appropriately by a Frenkel-Kontorova-like model, with the microscopic degrees of freedom now ``kinks''; the collective effect of an array of kinks is to introduce a slope to the domain wall on average.
The strain-mediated coupling will generate a long-range repulsive interaction between the kinks, and the {\em linear} response of the wall dynamics will at long-length scales reproduce the features we discuss here.
The dynamics and kinetics on short length scales are inherently non-linear, and different.
However, the gap for kink formation is small enough that near room temperature, there will be a proliferation of kinks and so our coarse-grained analysis of domain wall dynamics becomes valid.\cite{meyer_ab_2002,shin_nucleation_2007}

\section{Acknowledgments}
\label{sec:acknowledgements}

We acknowledge helpful discussions with A.~Saxena, T.~Lookman, and J.~F.~Scott.
Work at Argonne was supported by Basic Energy Sciences, Office of Science, U.S. Department of Energy, under Contract No. DE-AC02-06CH11357. RTB is supported by Homerton College, Cambridge.

\bibliography{out}

\end{document}